\documentclass[AMA,STIX1COL]{WileyNJD-v2}

\articletype{Research Article}%

\received{23 November 2021}
\revised{1 August 2022}
\accepted{7 September 2022}

\usepackage{tabularx}
\usepackage{graphicx}
\usepackage{textcomp}
\usepackage{xcolor}
\usepackage{booktabs}
\usepackage{lipsum}

\newcommand\keyword{\bgroup\markoverwith
  {\textcolor{yellow}{\rule[-.5ex]{2pt}{2.5ex}}}\ULon}
  
\newcommand\blfootnote[1]{%
\begingroup
\renewcommand\thefootnote{}\footnote{#1}%
\addtocounter{footnote}{-1}%
\endgroup
}

\newcommand{\makecell}[2]{\begin{tabular}{@{}#1@{}}#2\end{tabular}}
\usepackage{xspace}
\newcommand{\tool}{TRRGen\xspace}

\begin{document}

\title{A Transformer-Based Approach for Improving App Review Response Generation}

\author[1]{Weizhe ZHANG§}

\author[2]{Wenchao GU§}

\author[1,3,4]{Cuiyun GAO*}

\author[2]{Michael R. LYU}

\authormark{ZHANG \textsc{et al}}

\address[1]{\orgdiv{School of Computer Science and Technology}, \orgname{Harbin Institute of Technology}, \orgaddress{\state{Shenzhen}, \country{China}}}

\address[2]{\orgdiv{Department of Computer Science and Engineering}, \orgname{The Chinese University of Hong Kong}, \orgaddress{\state{Hong Kong}, \country{China}}}

\address[3]{\orgdiv{Guangdong Provincial Key Laboratory of Novel Security Intelligence Technologies}, \orgaddress{ \country{China}}}

\address[4]{\orgdiv{Peng Cheng Laboratory}, \orgaddress{ \country{China}}} 

\corres{*Cuiyun GAO, School of Computer Science and Technology, Harbin Institute of Technology, Shenzhen, China \\ \email{gaocuiyun@hit.edu.cn}}


\abstract[Summary]{Mobile apps are becoming an integral part of people's daily life by providing various functionalities, such as messaging and gaming. App developers try their best to ensure user experience during app development and maintenance to improve the rating of their apps on app platforms and attract more user downloads. Previous studies indicated that responding to users' reviews tends to change their attitude towards the apps positively. Users who have been replied are more likely to update the given ratings. However, reading and responding to every user review is not an easy task for developers since it is common for popular apps to receive tons of reviews every day. Thus, automation tools for review replying are needed.

To address the need above, the paper introduces a Transformer-based approach, named TRRGen, to automatically generate responses to given user reviews. TRRGen extracts apps' categories, rating, and review text as the input features. By adapting a Transformer-based model, TRRGen can generate appropriate replies for new reviews. Comprehensive experiments and analysis on the real-world datasets indicate that the proposed approach can generate high-quality replies for users' reviews and significantly outperform current state-of-art approaches on the task. The manual validation results on the generated replies further demonstrate the effectiveness of the proposed approach.}

\keywords{App reviews, review response generation, Transformer}


\maketitle

\blfootnote{\raisebox{3pt}{\small{§ }}\footnotesize{The first two authors contribute equally to the work.}} 

\section{Introduction}\label{sec:intro}
Mobile application (app) is a kind of software designed for mobile devices.  With the increase in 
market occupancy of mobile devices, the number of app users grows drastically in recent years. Nowadays, the number of smartphone users worldwide hits 6.378 billions\footnote{https://www.statista.com/statistics/330695/number-of-smartphone-users-worldwide/}, indicating that there is a vast and fiercely competitive market for apps. Most apps are downloaded via mobile application platforms, such as Apple's App Store and Google Play Store. Almost all of these platforms allow users to comment and rate the downloaded apps. Meanwhile, the application platforms generally allow developers to reply to user reviews, after which the corresponding users will receive notification and could choose to update the reviews and ratings. Obviously, the user review function on the application platform is one of the most effective channels for communication between users and developers. Since the average user rating score concerns the ranking of the app on the platform, developers commonly try their best to improve the user experience.

According to the research from McIlroy et al. \cite{mcilroy2015worth}, 38.7\% of users update their rating after the response from developers. Hassan et al. \cite{hassan2018studying} observed that 34.1\% of apps developers they analyzed respond to at least one review, and this responding action is demonstrated to have a positive effect for rating update. For example, they found that the number of users who change their given ratings after receiving a developer reply is six times more than those who have not received any response. They also discovered that 34\% of mentioned problems in user reviews could be solved sorely by developer's replies instead of updating the apps. The above results indicate that the review-response mechanism can improve both user experience and app ratings.

Despite the benefit of the review-response mechanism, it is a heavy and laborious task for developers to manually respond to every user review due to the large and ever-increasing number of reviews received daily. It highlights the significance and necessity of automation for the replies of the users' reviews. This paper focuses on creating a review response generation model that can more accurately
generate corresponding responses to given reviews.

Review response generation can be analogue to the dialogue generation task in the
Natural Language Processing (NLP) field~\cite{wang2017steering, li2018syntactically}. Some of the approaches are already deployed
in practical products, such as the XiaoIce chat bot\cite{zhou2020design} from Microsoft. This work applies the Neural Machine Translation model (NMT). NMT models are also applied to different NLP tasks such as translation, data-to-text generation, sentiment classification, etc \cite{luong2015effective,cho2014learning,puduppully2019data,tang2016effective}. Gao et al. \cite{gao2019automating} firstly utilize an Recurrent Neural Network (RNN) based NMT model named RRGen to generate the response for app reviews. However, RNN-based models have proven less effective in natural language comprehension comparing with Transformer-based models~\cite{vaswani2017attention,DBLP:journals/jmlr/RaffelSRLNMZLL20}. Besides, in Gao et al.'s work~\cite{gao2019automating}, the usefulness of review features such as app category and review rating have been validated. For example, review ratings could be beneficial to capture the user sentiment, e.g., whether the user is complaining about the apps. But the integration of the features into Transformer has never been explored in the review response generation task.

In this paper, we propose a Transformer-based review response generation approach (TRRGen). We integrate the Transformer model with the app's category and user rating features to capture the semantics of review texts and user sentiment. Two real-world datasets are used to evaluate the proposed model. One dataset contains 293,778 pairs of review-response data from 58 different apps published on the Google Play Store. The other dataset was also from Google Play Store but was recently crawled. 
We select four baseline models for comparison in the experiment. According to the experiment results, TRRGen outperforms the baseline approaches by at least 26.1\% in terms of BLEU-4 score \cite{papineni2002bleu}. We also conducted a human evaluation of the generated response from both the proposed and baseline models in the experiment. The results indicate that TRRGen can generate more relevant and accurate responses.

Overall, we make the following contributions:
\begin{itemize}
    \item{We propose TRRGen, a Transformer-based approach to predict developer's responses towards given user reviews.}
    \item{We conduct both automatic evaluation and manual evaluation on the performance of the proposed models and baselines. The experimental results indicate that TRRGen outperforms the state-of-the-art model by at least 26.1\% in terms of BLEU-4 score.}
\end{itemize}

The rest of this paper is organized as follows. Section~\ref{sec:back} introduces the background knowledge about User-Developer Dialogue in our task and Transformer. Section~\ref{sec:method} introduces an overview of the proposed approach and the detailed design of the approach. Section~\ref{sec:setup} illustrates the experimental datasets, evaluation metrics, and implementation details. Section~\ref{sec:result} elaborates on the experimental results, including the results from the automatic evaluation and manual evaluation. Section~\ref{sec:threat} introduces the threats of validity in our experiments. Section~\ref{sec:literature} surveys the related work and Section~\ref{sec:conclusion} concludes our work.

\section{Background}\label{sec:back}
In this section, we introduce the background of the work, including the user-developer dialogue and one basic deep learning technique - Transformer.

\subsection{User-Developer Dialogue}

\begin{figure}[h]
\centering
\includegraphics[width=0.6\textwidth]{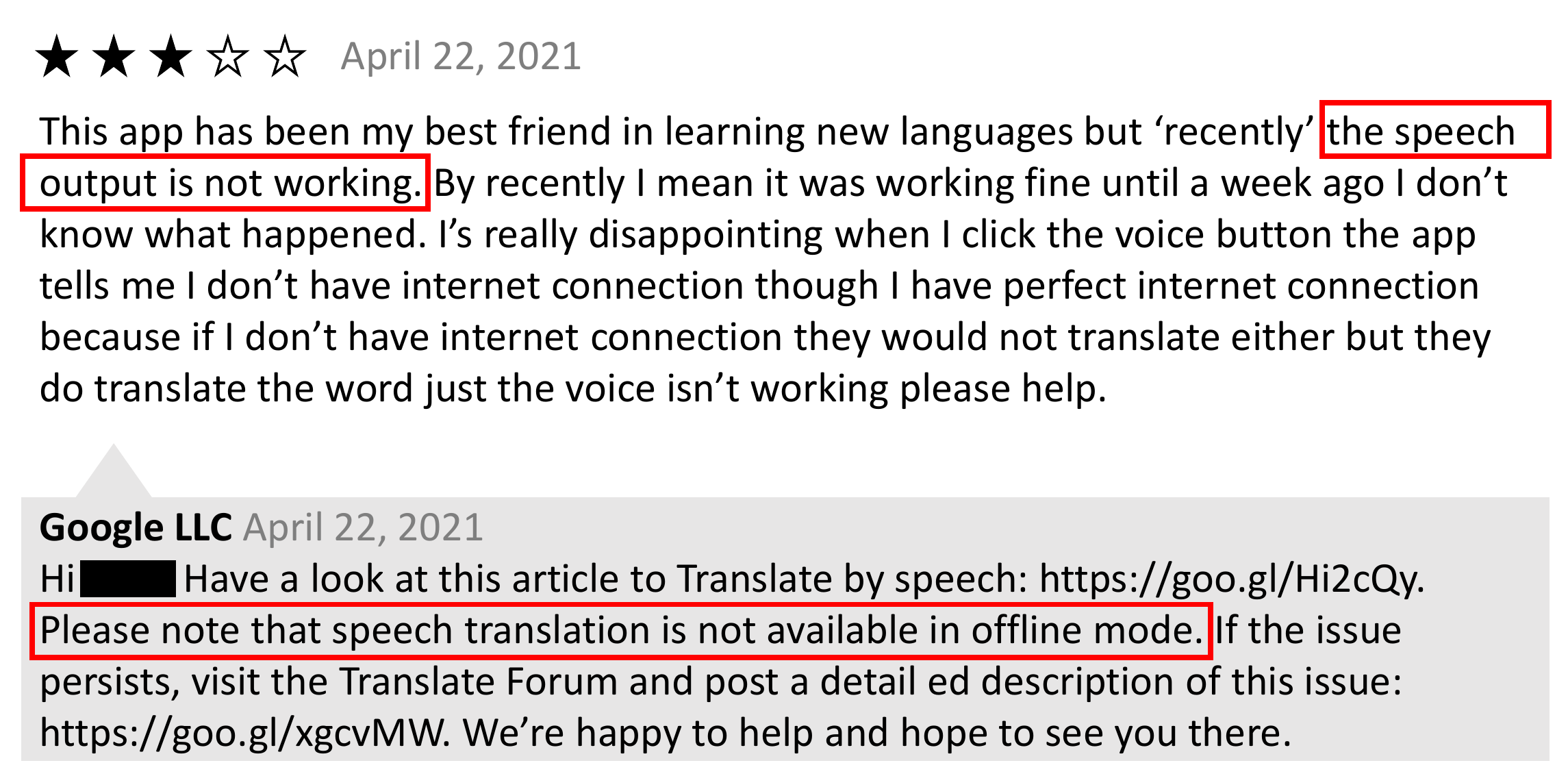}
\caption{Example of a developer's response to a user review. The words in red rectangle indicate the topic of the dialogue.}
\label{image1}
\end{figure}

Many mobile application platforms, including Google Play Store, allow users to download apps, after which users can post reviews and give a star rating to the apps. User reviews usually involve such topics: bug report, feature request, complaint, or praise about the experience \cite{maalej2015bug}. As for the star ratings, users can give one to five stars with their reviews to express whether they are satisfied with the user experience or willing to recommend the apps. Both of the given reviews and star ratings can be changed by users at any time. In order to receive a better rating and review, developers will try their best to enhance the user experience, and respond to users' reviews via the platforms is one important channel. Figure~\ref{image1} depicts an example of the user-developer dialogue from the Google Translate app. The user claimed that the ``speech output" function was not working correctly. Then the developer of Google Translate responded that the ``speech" function was not available in offline mode, which could probably solve the user's problem.

Since the number of historical and new reviews in popular apps is significantly huge, developers apparently cannot reply to all reviews. A dataset\footnote{https://www.kaggle.com/lava18/google-play-store-apps} indicates that 39.48\% of the collected apps receive over 10,000 reviews, and 20.00\% of them receive even over 100,000 reviews. However, as studied by Hassan et al. \cite{hassan2018studying}, developers only respond to 2.8\% of the collected reviews, and they tend to reply to the reviews with low ratings and long content. Moreover, 97.5\% of the user-developer dialogues end after one round. Thus, we focus on alleviating manual labor by automatically generating one iteration of the user review reply in our study.

\subsection{Transformer}
Transformer follows a typical encoder-decoder structure which is generally composed of $N$ stacked Transformer blocks. In the Transformer encoder, each block contains a multi-head self-attention sub-layer and a fully connected positional-wise feed-forward network sub-layer. The Transformer decoder has a similar composition as the encoder. Besides the mentioned sub-layers above, a masked multi-head self-attention sub-layer ensures that the predictions for position $i$ can depend only on the known outputs at positions less than $i$. Each sub-layer is connected by the residual connections \cite{he2016deep} and layer normalization \cite{ba2016layer}. To encode word position information, the Transformer calculates each input word's positional tensor using sine and cosine functions of different frequencies and adds them to the input embedding. Next, we introduce positional encoding, multi-head self-attention, position-wise feed-forward networks, and the process of residual connections and layer normalization.

\subsubsection{Positional Encoding}
RNN-based and CNN-based models can naturally learn the order of the sequence. However, the self-attention mechanism does not have such a feature. To inject the information about the relative or absolute position of the tokens in the sequence, Transformer augments a process of positional encoding with token embedding. The positional encoding has the same dimension as the embedding, so its output can be summed. The Transformer uses sine and cosine functions of different frequencies in a positional encoding:

\begin{equation}
    PE_{(pos,2i)} = {\rm sin}(pos/10000^{2i/d_{model}})
\end{equation}
\begin{equation}
    PE_{(pos,2i+1)} = {\rm cos}(pos/10000^{2i/d_{model}})
\end{equation}

where $pos$ is position of the word in sequence and $i$ is the order of the dimension.

\subsubsection{Multi-head Self-Attention}
The self-attention mechanism can be described as mapping a query and a set of key-value pairs to an output:
\begin{equation}
    attention_{output} = {\rm Attention}(Q,K,V)
\end{equation}
where $Q$, $K$, $V$ is the vector of query, key and value respectively. 

\textbf{Scaled Dot-Product Attention.} The self-attention used in Transformer is also known as scaled dot-product attention. Transformer first projects the input vector $X$ into query vector $Q$, key vector $K$, and value vector $V$ by multiplying it with trainable parameters $W^{Q}$, $W^{K}$ and $W^{V}$. The attention weight is calculated using dot product and softmax function:
\begin{equation}
   {\rm Attention(Q,K,V)} = {\rm softmax}(\frac{QK^T}{\sqrt{d_k}})V
\end{equation}
The output of the attention function contains the information of relation between every word in the input sequence.
A multi-head self-attention sub-layer has $h$ attention heads and applies self-attention mechanism on each head since it is demonstrated to be more effective to perform the attention function in parallel. One attention head obtains one representation space for the same text, and multi-head attention obtains multiple representation spaces. Then the information from different representation sub-spaces will be combined into output $Z$. In other words, multi-head attention captures different contexts with multiple individual self-attention functions. Multi-head attention is computed after scaled dot-product attention:
\begin{equation}
    head_{i} = {\rm Attention}(QW_{i}^{Q},KW_{i}^{K},VW_{i}^{V})
\end{equation}
\begin{equation}
    Z = {\rm MultiHead}(Q,K,V) = {\rm Concat}(head_{1}...head_{h})W^{O}
\end{equation}
where parameters $W_{i}^{Q}$, $W_{i}^{K}$, $W_{i}^{V}$ are independent in each head, and $W^{O}$ is the trainable weight matrix of multi-head attention sub-layer.

\subsubsection{Position-wise Feed-Forward Networks}
A fully connected feed-forward network (FFN) sub-layer is after the multi-head attention sub-layer in each block of encoder and decoder. This FFN can be computed by two linear transformations and a ReLU activation between them.
\begin{equation}
    {\rm FFN(x)} = {\rm max}(0,xW_{1}+b_{1})W_{2}+b_{2}
\end{equation}
where $W_{1}$, $W_{2}$ are weight matrices and $b_{1}$, $b_{2}$ are bias matrices.

\subsubsection{Residual Connections and Layer Normalization}
The residual connections and layer normalization are conducted in every sub-layer of each encoder or decoder block. Residual connections solve the vanishing gradient problem, while layer normalization avoids excessive deviation of data after multi-layer calculation. This process happens like this:
\begin{equation}
\label{eq1}
    L = {\rm LayerNorm}(X + Z)
\end{equation}
\begin{equation}
\label{eq2}
    S = {\rm LayerNorm}(L + FNN(L)) 
\end{equation}
where $LayerNorm$ is the normalization function, $X$ and $Z$ are input and output of multi-head attention sub-layer respectively. Equ. (\ref{eq1}) is computed after multi-head attention sub-layer and Equ. (\ref{eq2}) is computed after FNN.

\section{Methodology}\label{sec:method}

\begin{figure*}[h]
\centering
\includegraphics[width=1\textwidth]{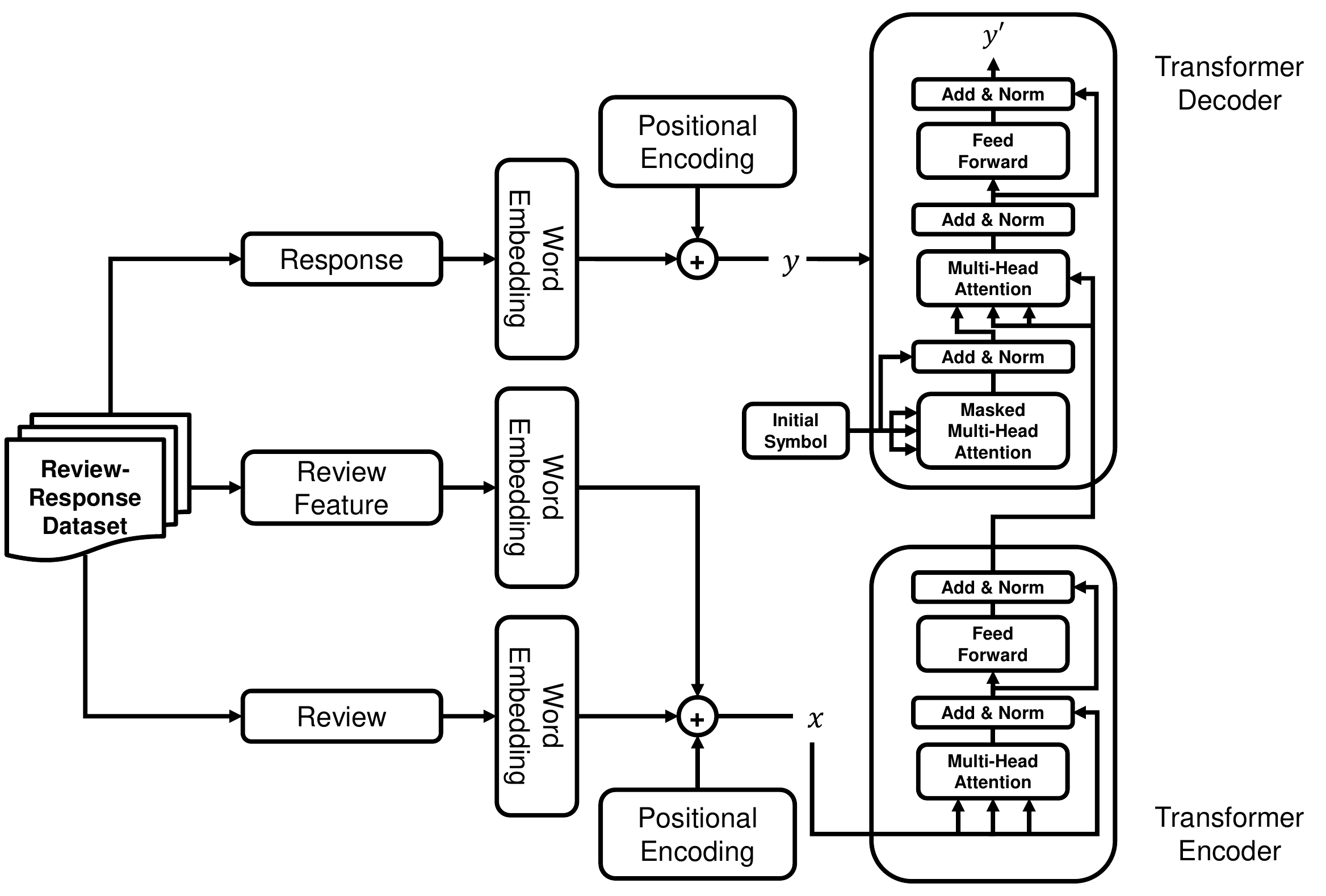}
\caption{Architecture of the proposed Transformer-based review response generation model.}
\label{image2}
\end{figure*}

We propose to adopt Transformer \cite{vaswani2017attention} to generate the response to the given user review. Fig. \ref{image2} is the overall structure of our model. Both user review and response will be extracted as a sequence of tokens and then represented as a sequence of vectors, which are $\textbf{x}=(x_1,...,x_n)$. This section introduces how to model the tokens with app categories, rating, and their position in user reviews.

\subsection{User Review Embedding}

\textbf{Rating Feature Embedding:}
Developers will adopt different strategies to respond to user reviews according to the user emotion hidden in the reviews. However, such emotion will not be explicitly written in the review concept, but we have the rating from users. Rating plays an important role in quantifying review sentiment since it is positively correlated with the user's emotion intuitively (User's review will have a positive sentiment with a high rating if the user feels satisfied about the app, and vice versa). Since the digits of the rating can be mixed up with the numbers in the text, we first replace the rating with escape characters. For example, replace ``4'' with ``$\langle4\rangle$''. Then it goes through the process of word embedding and becomes a rating feature vector. Finally, we add the rating feature vector to each token vector in the corresponding sequence:
\begin{equation}\label{equ:rating}
    x_i = w_i + r + p_i,
\end{equation}
where $w_i$ is the $i^{th}$ vector in the input review, $r$ is the rating feature vector and $p_i$ is the output of positional encoding. To acquire the word embedding vector, we adopt a trained embedding layer to convert the one-hot vector into a dense embedding vector. Adding the rating feature vector into the token embedding vectors can enhance the emotional information for the token. Such fusion enables the encoder to explicitly learn the emotion hidden in user reviews and thereby the decoder can generate responses with more appropriate emotion.

\begin{figure}[h]
\centering
\includegraphics[width=0.5\textwidth]{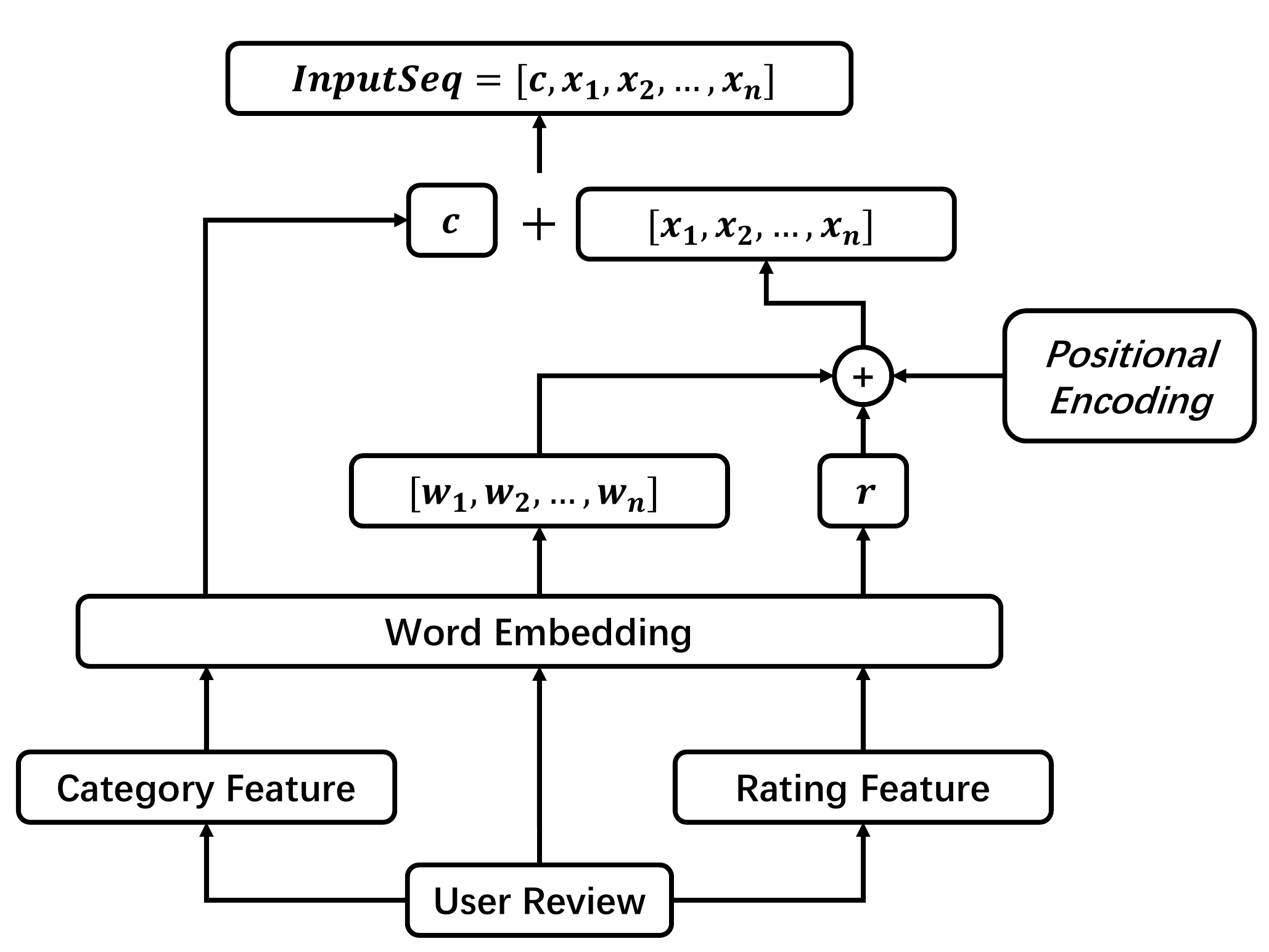}
\caption{Illustration of the process of embedding user reviews in our model.}
\label{image3}
\end{figure}

\textbf{App Category Feature Embedding:}
Developers of the same app categories usually respond in similar topics. To learn the knowledge of the topics for same categories, the model should provide the information about the app categories associated with given reviews. Thus we concatenate the app category and the intermediate input sequence from Equ.~(\ref{equ:rating}):
\begin{equation}
    X = [c,x_1,x_2,...,x_n]^T
\end{equation}
where $c$ is the category feature vector, $X$ is both the output sequence of user review embedding and the input sequence of the encoder. Fig. \ref{image3} describes how we transform review text and corresponding feature tokens into the ultimate input sequence. Since the self-attention mechanism in Transformer is helpful for learning the long-term dependency relationship~\cite{vaswani2017attention}, the information inside the category embedding vectors can still be well retained in the last layer of the encoder and passed to the decoder. The decoder then is able to generate the responses with different styles according to this category information.    

\section{Experimental Setup}\label{sec:setup}

This section mainly introduces the datasets, data preprocessing, and evaluation metrics we have used during the experiments. We also introduce three baseline approaches for comparison.

\subsection{Datasets Description}
In our experiments, we have used two different datasets, denoted as dataset A and dataset B, respectively. 

Dataset A is a benchmark dataset proposed by Gao et al.
\cite{gao2019automating}, collected from Google Play Store from 2016 to 2018. Dataset A contains 293,778 pairs of review-response data. Specifically, it contains 278,374 pairs of data in the training set, 7,702 pairs of data in the validation set, and 7,702 pairs of data in the test. The data were gathered from 58 different apps, which respectively contain at least 100 user reviews according to the statistics. 


\begin{table}[htbp]
\caption{The average numbers of word of review and response in dataset A and dataset B}
\label{table2}
\centering
\setlength{\tabcolsep}{2.5mm}
\begin{tabular}{ccc}
\toprule
 & review text & response text  \\
\midrule
Dataset A & 16.03 & 48.90 \\
Dataset B & 30.22 & 32.68 \\
\bottomrule
\end{tabular}
\end{table}  

Dataset B was collected by ourselves in the early of 2021. It contains 110,832 pairs of review-response data from 18 apps on the Google Play Store. Dataset B has the same data source as dataset A. We assign 5,000 pairs of data as the test set and the rest of them as the training set. Table \ref{table2} describes the average numbers of words about review text and response text in both dataset A and dataset B. In dataset B, the average numbers of words in review texts and response texts are 30.22 and 32.68, respectively, which are largely different from those of dataset A. It seems users now prefer leaving longer reviews, but developers reply more succinctly. The differences in the collection dates and lengths of review-response text indicate that it is meaningful to validate our model on both datasets. 

\subsection{Data Preprocessing}
Some advertisement-like sentences in the response set are meaningless for training our model, so these sentences should be removed. We have observed the response text and found that most ad sentences exist in the middle of the whole response text. The following sentences are two examples we found in our dataset:

\textit{``$\langle app\_name\rangle$ is a free phone cleaner which commit to scan virus and clean virus, do the best protection, clean cache, clean junk file, boost phone speed, save battery and cool down phone temperature for your android phone.''}

\textit{``Sorry for the inconvenience caused, we'll keep improving $\langle app\_name\rangle$ to provide users with better experience.Thanks for your support.''}

Above template-based replies do not provide any useful information to the users, and generating such replies is meaningless for both developers and users, thereby we remove them from the original data. To remove these ad sentences, we selected five continuous words as an expression in the middle of each sentence, then counted the frequency that the expression exists in the whole response corpus. We manually judge if the most frequent expression belonged to an ad sentence and determined if it should be removed.

There are also a large number and different kinds of specific words like user ID, email address, app's name, and URL in the response corpus. They would occupy much space in the dictionary and could cause an OOV (Out-of-Vocabulary) problem since they are usually unique for every app. Thus, we used several regular expressions to detect these words and replace them with escape characters. For example, we replace all email addresses, app's names, user IDs and URLs by ``$\langle email\rangle$'', ``$\langle app\_name\rangle$'', ``$\langle user\_name\rangle$'', and ``$\langle url\rangle$'', respectively.

\subsection{Evaluation method - BLEU}
BLEU \cite{papineni2002bleu} (Bilingual Evaluation Understudy) was initially designed to evaluate the effectiveness of machine translation, and now it is also considered as a standard automatic metric for evaluating dialogue response generation systems. In particular, BLEU is used to evaluate the difference between candidate sentences (generated by machine) and reference sentences, representing a score in a range of 0 to 1. In our cases, candidate sentences are model-generated responses, and reference sentences are real developer's replies. 
BLEU analyzes the co-occurrences of n-grams in the real replies $y$ and the generated responses $\hat{y}$, where n can be 1, 2, 3, and 4. Here is the formula for calculating n-gram accuracy:
\begin{equation}
    p_n = \frac{\sum_{c\in candidates}\sum_{n-gram\in c} Count_{clip}(n-gram)}{\sum_{c^{'}\in candidates}\sum_{n-gram^{'}\in c^{'}} Count(n-gram^{'})}
\end{equation}
where $c$ or $c^{'}$ represents a sentence selected from the candidate set, $Count_{clip}(n-gram)$ is the number of the n-gram existing in the corresponding reference sentences, and $Count(n-gram^{'})$ counts the number of the $n-gram^{'}$ existing in candidate sentences. So the whole molecule is how many n-gram words in a given candidate appear in the reference, while the denominator is the number of n-grams in all candidates.

For those $\hat{y}$ shorter than $y$, BLEU reduces their score by multiplying a Brevity Penalty factor because shorter candidate sentences may not fully express the meaning of reference sentences. In BLEU-N, N means the maximum length of n-grams considered, measures the proportion of co-occurrences of n consecutive tokens between $y$ and $\hat{y}$:
\begin{equation}
    BLEU = BP\cdot\exp{\sum_{n=1}^{N}w_{n}\log p_{n}}
\end{equation}
where $w_n = 1/N$. Generally, we use the $N=4$ version of BLEU, that is, BLEU-4. BLEU-4 is a weighted geometric mean of $p_1$ to $p_4$ (in our case, the weights are (0.25, 0.25, 0.25, 0.25). BLEU-4 is often used for corpus-level evaluation, which is more correlated with human judgments than other evaluation metrics~\cite{gao2019automating}.

\subsection{Baseline Approaches}
We compare the performance of our model with the random selection approach, vanilla Transformer model \cite{vaswani2017attention} and RRGen model\cite{gao2019automating}. 

\textbf{Random selection:} This is a strawman baseline. This approach randomly chooses a response in the training set as the predicted result to a given review in the test set.

\textbf{Vanilla Transformer:} This is the vanilla Transformer model with the same size as the Transformer we used in our model. We directly use user reviews as input and developer's replies as output to train the Transformer. This transformer model is used in ablation experiments.

\textbf{RRGen:} RRGen adopts a different deep learning model for review response generation. RRGen leverages a Bi-LSTM network \cite{huang2015bidirectional} as its encoder, applies an attention model to process the output of the encoder, and then uses an LSTM network to decode output. In the attention model, RRGen combines the attention information with the review features including user ratings and app categories, review sentiment, and keywords.

\section{Experimental Results}\label{sec:result}

This section presents the evaluation results of the proposed model, including main results, parameters analysis, ablation study, case studies, and human evaluation.

\subsection{Main Results}

\begin{table}[t]
\centering
\caption{Comparison results with baseline models on the Dataset A. The $p_n$ indicates the $n$-gram precision when comparing the ground truth and generated responses. The best results are highlighted in \textbf{bold} fonts.}
\label{tab:main_result_a}
\begin{tabular}{cccccc}
\toprule
Approach & BLEU-4 & $p_1$ & $p_2$ & $p_3$ & $p_4$  \\
\midrule
Random Selection & 3.12 & 19.62 & 3.93 & 1.36 & 0.916\\
Vanilla Transformer & 33.12 & 50.66 & 32.72 & 28.05 & 25.87\\
RRGen & 35.99 & 51.64 & 35.33 & 30.91 & 29.75 \\
\midrule
TRRGen & \textbf{45.38} & \textbf{60.81} & \textbf{45.07} & \textbf{40.55} & \textbf{38.15} \\
\bottomrule
\end{tabular}
\end{table}

/*\begin{table}[t]
\centering
\caption{Comparison results with baseline models on the Dataset B. The best results are highlighted in \textbf{bold} fonts.}
\label{tab:main_result_b}
\begin{tabular}{cccccc}
\toprule
Approach & BLEU-4 & $p_1$ & $p_2$ & $p_3$ & $p_4$  \\
\midrule
Random Selection & 2.56 & 15.00 & 2.56 & 1.22 & 0.92\\
Vanilla Transformer & 24.51 & 40.12 & 26.70 & 23.30 & 21.69\\
RRGen & 30.77 & 48.01 & 33.17 & 29.54 & 27.97 \\
\midrule
TRRGen & \textbf{36.08} & \textbf{49.95} & \textbf{37.30} & \textbf{33.64} & \textbf{31.65} \\
\bottomrule
\end{tabular}
\end{table}

\textbf{The fusion of rating and app category features can improve the performance of response generation}. Table~\ref{tab:main_result_a} and Table~\ref{tab:main_result_b} illustrate the evaluation results comparing with the baseline models. Both RRGen and TRRGen, which fuse the review features into the models, outperform the vanilla Transformer that only utilizes the tokens in the review text. TRRGen adopts the same model architecture as vanilla Transformer but increases the performance of 37.02\% in terms of the BLEU-4 metrics, 20.04\% in terms of $p_1$, 37.74\% in terms of $p_2$, 44.56\% in terms of $p_3$ and 47.47\% in terms of $p_4$ in the dataset A. TRRGen also outperforms the vanilla Transformer 47.21\%, 24.50\%, 39.70\%, 44.38\% and 45.92\% in terms of BLEU-4, $p_1$, $p_2$, $p_3$ and $p_4$, respectively.  Besides, the performance of RRGen is much worse than the performance of TRRGen. However, RRGen still outperforms vanilla Transformer in all metrics with a low-accuracy deep learning model. These results demonstrate the effectiveness of fusing rating and app category features into deep learning models. 

\textbf{Self-Attention mechanism help our proposed model generate higher-quality responses}. As can be seen in Table~\ref{tab:main_result_a}, TRRGen outperforms RRGen 26.09\% in terms of BLEU-4, 17.76\% in terms of $p_1$, 27.57\% in terms of $p_2$, 31.19\% in terms of $p_3$ and 28.24\% in terms of $p_4$ in the dataset A. According to Table~\ref{tab:main_result_b}, TRRGen outperforms RRGen 17.26\%, 4.04\%, 12.45\%, 13.88\% and 13.16\% in terms of BLEU-4, $p_1$, $p_2$, $p_3$ and $p_4$, respectively.  The cause of the performance improvement is the self-attention mechanism. Although the design of Bi-LSTM is to capture long-term dependence information,  long-term dependence information is still inevitably lost when the input is quite long. As demonstrated in previous studies in natural language processing~\cite{vaswani2017attention,DBLP:journals/jmlr/RaffelSRLNMZLL20}, the self-attention mechanism in the Transformers can help the model capture the long-term dependence information better. The self-attention mechanism also works in our proposed model under the task of response generation.

\subsection{Parameters Analysis}

In this section, we will discuss how hyper-parameters affect the performance of TRRGen. Two hyper-parameters are analyzed: the number of multi-head attention and the number of layers in both encoder and decoder.

\begin{table}[t]
\centering
\caption{Evaluation results with different numbers of multi-head attention. The best results are highlighted in \textbf{bold} fonts.}
\label{tab:multi_head_number}
\begin{tabular}{cccccc}
\toprule
\makecell{c}{Number of\\Multi-Head Attention} & BLEU-4 & $p_1$ & $p_2$ & $p_3$ & $p_4$  \\
\midrule
1 & 44.23 & 59.83 & 43.86 & 39.40 & 37.02 \\
2 & 43.42 & 58.98 & 43.05 & 38.60 & 36.29 \\
4 & \textbf{45.38} & 60.81 & \textbf{45.07} & 40.55 & 38.15 \\
8 & 44.87 & \textbf{60.90} & 44.98 & \textbf{40.56} & \textbf{38.22} \\
\bottomrule
\end{tabular}
\end{table}

\begin{table}[t]
\centering
\caption{Evaluation results with different layer number. The best results are highlighted in \textbf{bold} fonts.}
\label{tab:layer_number}
\begin{tabular}{cccccc}
\toprule
{Layer Number} & BLEU-4 & $p_1$ & $p_2$ & $p_3$ & $p_4$  \\
\midrule
1 & \textbf{45.38} & \textbf{60.81} & \textbf{45.07} & \textbf{40.55} & \textbf{38.15} \\
2 & 43.56 & 59.11 & 43.16 & 38.76 & 36.42 \\
3 & 40.41 & 56.01 & 40.00 & 35.66 & 33.39 \\
4 & 39.22 & 55.33 & 38.54 & 34.36 & 32.30 \\
5 & 38.91 & 54.85 & 38.39 & 34.11 & 31.91 \\
6 & 38.06 & 54.55 & 37.38 & 33.25 & 31.36 \\
\bottomrule
\end{tabular}
\end{table}

\subsubsection{Number of Multi-Head Attention}

As shown in Table~\ref{tab:multi_head_number}, the performance of the proposed model in all metrics has an increasing trend as the number of multi-head attention grows, although the tendency is not so apparent. It is easy to understand since the multi-head attention mechanism allows the model to have multiple sets of attention scores on the input to understand the input from different aspects. We can observe that the performance of the proposed model is very close when the number of multi-head attention is defined as 4 or 8. Considering the computation efficiency, we define the number of multi-head attention as 4 in our experiment.

\subsubsection{Number of Layer in Both Encoder and Decoder}

As shown in Table~\ref{tab:layer_number}, the performance of the proposed model in all metrics drops with the increase of layer number in both encoder and decoder. Overfitting is the leading cause of this phenomenon. The training data size may not be large enough so that a complex model may learn noise or fake features only appeared in the training data. We choose 1 as the layer number in both encoder and decoder in our experiment to avoid such overfitting.

\begin{table}[t]
\centering
\caption{Ablation Study on Dataset A. The best results are highlighted in \textbf{bold} fonts.}
\label{tab:ablation_study}
\begin{tabular}{cccccc}
\toprule
Approach & BLEU-4 & $p_1$ & $p_2$ & $p_3$ & $p_4$  \\
\midrule
Vanilla Transformer & 33.12 & 50.66 & 32.72 & 28.05 & 25.87 \\ 
{$\rm Transformer_{Rating}$} & 30.87 & 48.23 & 30.46 & 25.94 & 23.83 \\
{$\rm Transformer_{Category}$} & 43.45 & 59.04 & 43.02 & 38.63 & 36.31 \\
TRRGen & \textbf{45.38} & \textbf{60.81} & \textbf{45.07} & \textbf{40.55} & \textbf{38.15} \\
\bottomrule
\end{tabular}
\end{table}

\subsection{Ablation Study}

In this section, we validate the contribution of the rating feature or app category feature and the effectiveness of fuse both features. Table~\ref{tab:ablation_study} shows the results of the ablation study on dataset A. Vanilla Transformer represents the model that only utilizes the review text. {$\rm Transformer_{Rating}$} represents the model that fuses the review text with the rating feature. {$\rm Transformer_{Category}$} represents the model that combines the review text with the app category feature. TRRGen represents the model that fuses the review text with both rating feature and app category feature. 

From the result, we can find that {$\rm Transformer_{Category}$} outperforms vanilla Transformer 31.19\% in terms of BLEU-4, 20.04\% in terms of $p_1$, 37.74\% in terms of $p_2$, 44.56\% in terms of $p_3$ and 47.47\% in terms of $p_4$. This result demonstrates that involving app category features can greatly improve the quality of generated responses. We can also observe that the performance of {$\rm Transformer_{Rating}$} is dropped compared to the performance of vanilla Transformer, which indicates that simply fusing the user review with the rating feature will disturb the model to generate the proper response. However, there is an interesting phenomenon that TRRGen outperforms all other models. This result indicates that the rating feature will not disturb the model anymore and can contribute to the performance improvement when the rating feature is fused with the app category feature. From the above results, we can conclude that the combination of app category feature and rating feature is effective.

\begin{table*}[t]
\centering
\caption{Successful Case 1. Keywords are marked with yellow background color. Suggestions are marked with underlines.}
\label{tab:example1}
\begin{tabular}{c|l}
\toprule
User’s review & \makecell{l}{You tell me to send u screenshots of the \keyword{rude news article} so u could remove them  and I do and u say you \\ would check again for the quality of the news. And today I have seen \keyword{2 new rude news articles}, this is \\ unacceptable. Can u sort this out or I will uninstall.} \\
\midrule
Real reply & \makecell{l}{Hi, we have rechecked \keyword{news problem} once we receive your feedback. There should be no problem on your \\ side. \underline{Could you send us screenshot of the} \keyword{2 rude news articles} \underline{via {$\rm \langle url\rangle$} again?} We will deal with the \\ problem. Thank you.} \\
\midrule
RRGen & \makecell{l}{Hi. Thank you for use {$\rm \langle appname\rangle$}. {$\rm \langle appname\rangle$} is a free phone cleaner which commit to scan virus and \\ clean master do you dislike? Feel free to contact us via {$\rm \langle url\rangle$}. Thanks.} \\
\midrule
TRRGen & \makecell{l}{Dear user, sorry for the inconvenience and we are trying to optimize the \keyword{news content} on charge screen. \\ \underline{Could you tell us more detail about this problem? Please send us more detail via {$\rm \langle url\rangle$}}. thanks!} \\
\bottomrule
\end{tabular}
\end{table*}

\begin{table*}[t]
\centering
\caption{Successful Case 2. Keywords are marked with yellow background color. Suggestions are marked with underlines.}
\label{tab:example2}
\begin{tabular}{c|l}
\toprule
User’s review & It \keyword{runs too slowly}! \\
\midrule
Real reply & \makecell{l}{Hey there \underline{please make sure that you are connecting to} \keyword{fast internet} \underline{and try again.} If the issue remains you \\ can always contact us at {$\rm \langle email\rangle$}! We would love to help you out from there.} \\
\midrule
RRGen & Hey there thanks for your feedback. \\
\midrule
TRRGen & \makecell{l}{Hi there, thanks for the review. We’re very sorry to hear that the \keyword{app run slowly} on your device. Please take \\ the following step to have the issue resolve. \underline{Go to your device setting application, manager, {$\rm \langle appname\rangle$}}, \\ \underline{clear data.} If the issue still continue please contact us at {$\rm \langle email\rangle$} we will try to help you.} \\
\bottomrule
\end{tabular}
\end{table*}

\subsection{Case Studies}

This section will discuss two examples from dataset A, including user's review, real reply, and reply generated by RRGen and TRRGen. In the examples, all the keywords are marked with yellow background color, and all the suggestions are marked with underlines.

In the example shown in Table~\ref{tab:example1}, the user complains that he has already reported the problem about rude news articles to the developers but still received two new rude news articles. Developers mention that they have rechecked the problem and ask the user to send the screenshot of the two rude news articles to them. We can observe that the reply generated by RRGen misunderstood the user's sentiment and failed to capture the correct keywords in the user's review. Due to the above mistakes, RRGen generates an irrelevant reply to this user's review. However, TRRGen noticed the anger from the user and made an apology for the problem. Besides, TRRGen identifies the keywords ``news content" and offers a similar suggestion as the developer's real reply to users.

In the example shown in Table~\ref{tab:example2}, the user complains that the app runs too slowly, and developers suggested the user switch to faster internet and send them an email if the issue remains. Similarly, as in the previous example, the reply generated by RRGen did not understand the user's intention and is irrelevant to the user's review. Nevertheless, TRRGen correctly caught the problem that the app ran slowly and apologized for it. Besides, the reply generated by TRRGen also offered the user a suggestion to clear the app data and ask the user sent them an email if the issue remains, which is more natural. Although the solution provided by \tool is not the same as the solution from the developers, both of them are possible solutions to deal with the “slow problem” in this case. So the suggestion provided by \tool is still reasonable.

\begin{table*}[t]
\centering
\caption{Failure Case. Keywords are marked with yellow background color. Grammar errors or unclear sentences are marked with underlines}
\label{tab:failed_example}
\begin{tabular}{c|l}
\toprule
User’s review & \makecell{l}{The advertisement recently pretty disturbs my work. I don't mind if I need to pay to \keyword{remove the ads}. thx.} \\
\midrule
Real reply & \makecell{l}{Hi there thank for your review please note that the ads be essential for us for fund so that we can continue \\ update and improve the app for our user however I will pass your suggestion regard remove the ads to our \\ 
developer to be consider if you have more question don't hesitate to contact us on {$\rm \langle email\rangle$}.} \\
\midrule
RRGen & \makecell{l}{Hey we are sorry to hear that you \underline{be have} issue with {$\rm \langle appname\rangle$}. Please contact us at {$\rm \langle email\rangle$} explain the \\ issue and we will help you out from there.} \\
\midrule
TRRGen & \makecell{l}{Hey there thank for your review we be sorry to hear that you be disappoint with our app. \underline{We would like to} \\ \underline{inform you that we have some hard to provide the ads you face an issue with {$\rm \langle appname\rangle$} gold subscription} \\ \underline{which will be the ads} and we will \keyword{give you the} \keyword{possibility to control the ads} and we apologize for that. If \\ you have any question please contact us at {$\rm \langle email\rangle$}.} \\
\bottomrule
\end{tabular}
\end{table*}

\subsection{Error Analysis}
This section will discuss one failure example from dataset A, including user's review, real reply, and reply generated by RRGen and TRRGen. In the example, all the keywords are marked with yellow background color, and all the grammar errors or unclear sentences are marked with underlines.

In the example shown in Table~\ref{tab:failed_example}, the user states that the advertisement disturbs his work, and he is willing to pay to remove the ads. The importance of ads to the app is explained in the real reply, and it also mentioned that the user's suggestion would be passed to the developers. RRGen seemed not to understand the problem and generated a template-like reply. Besides, the reply generated by RRGen contains a grammar error ``be have". TRRGen caught the keyword ``ads" and replied to the user that they would give him the possibility to control ads, which can be a reasonable solution. However, TRRGen generated quite a long sentence that contains grammar errors and cannot be understood by the user. It demonstrates that the proposed model still can generate a reply with very low quality, and the reason for such phenomenon is also tough to explain due to the poor interpretability of the current deep learning model. 

\subsection{Human Evaluation}

This section will discuss the human evaluation result of TRRGen, RRGen, and real responses in reality. The human study can accurately evaluate user's satisfaction with the given responses, while the automatic measurement methods such as BLEU only measure the textual similarity between the generated responses and ground truth. In the experiment of human evaluation, we invite six participants for our survey, including three Ph.D. students and three master students, five of them have experience in software development for at least a year. 

In the survey, we randomly made 50 different review-response pairs from dataset A into a detailed questionnaire, but it will take a lot of time to complete all of them by one annotator at once. To reduce the workload of annotators, we follow the work~\cite{gao2019automating,GaoZXLXL22} to only put 25 examples in one questionnaire. The examples in each questionnaire are randomly selected from the total examples. During human evaluation, the two questionnaires are assigned with the same number of annotators. The annotators are asked to read 25 user reviews and the corresponding responses generated by TRRGen, RRGen, and developers. Then, they evaluate each response in three aspects, which are ``grammatical fluency", ``relevance", and ``accuracy", inspired by \cite{gao2019automating}. These evaluation metrics are described at the beginning of each questionnaire as below: 
\begin{itemize}
\item[*] The ``grammatical fluency" measures the degree of whether a text is easy to understand. 
\item[*] The ``relevance" relates to the extent of the topic similarity between the user review and response. 
\item[*] The ``accuracy" estimates the degree of sentiment consistency between the user review and response. For example, a negative review should not be responded in a grateful way.
\end{itemize}

These three metrics are rated on a 1-5 scale (5 for fully satisfying the rating scheme, 1 for completely not satisfying the rating scheme). Besides the three metrics, each participant is asked to rank the three given responses to the same user reviews based on their preference. The ``preference rank" score is rated on a 1-3 scale (1 for the most preferred).

\begin{table}[t]
\centering
\caption{Statistics of successful cases and failure cases in human evaluation.}
\label{tab:statistics}
\begin{tabular}{cccc}
\toprule
& \tool & Developer & RRGen  \\
\midrule
Successful cases & 40 & 43 & 28\\
Failure cases & 10 & 7 & 22\\
\bottomrule
\end{tabular}
\end{table}

\begin{table}[t]
\centering
\caption{Human evaluation result. The best results are highlighted in \textbf{bold} fonts.}
\label{tab:human_evaluation_result}
\begin{tabular}{ccccc}
\toprule
 & \makecell{c}{Grammatical\\Fluency} & Relevance & Accuracy & Preference Rank \\
\midrule
RRGen & 4.467 & 3.093 & 3.033 & 2.387 \\
TRRGen & \textbf{4.553} & 3.847 & 3.707 & 1.953 \\
\midrule
Developer & 4.547 & \textbf{4.040} & \textbf{3.887} & \textbf{1.66} \\
\bottomrule
\end{tabular}
\end{table}

Table~\ref{tab:statistics} shows the statistics of successful cases and failure cases in human evaluation from \tool, developer, and RRGen, respectively. We regard the reply whose scores for all the metrics of grammatical fluency, relevance, and accuracy are 4 or higher as the successful case, otherwise as the failure case. As shown in Table~\ref{tab:statistics}, only around half of the replies generated by RRGen can be regarded as successful cases while 80\% of replies generated by \tool can be regarded as successful cases, which demonstrates that the quality of generated replies from \tool is much higher than the previous baseline. In addition, it should be noticed that fewer than 90\% of replies can be regarded as successful cases even if the replies are written by developers since not all the replies are attentively written by developers. This result demonstrates that the quality of replies generated by \tool is very close to the quality of replies written by developers.

Table~\ref{tab:human_evaluation_result} shows the result of human evaluation. Bold indicates the top score in each metric. We can see that the response from developers is preferred over the other two approaches' outputs, which is within our expectation. Specifically, the responses from developers get the highest average score in the ``relevance" and ``accuracy" metrics, meaning that participants consider the developers' responses are the most relevant and sentiment consistent with the corresponding user reviews. However, TRRGen also does a good job. It achieves scores that are rather close to those of developers' responses. In terms of the ``Grammatical Fluency" metric, all approaches perform quite well, and their sores are significantly close, TRRGen even outperforms human developers in this metric. In addition, we can see that our TRRGen approach performs better across all metrics than the baseline approach, which also indicates that the improvement of TRRGen in review response generation.

\section{Discussion}
In this section, we attempt two different strategies for combing category information with semantic information, which are named as $\rm TRRGen_{\rm Sum}$ and $\rm TRRGen_{\rm Order}$, respectively. For $\rm TRRGen_{\rm Sum}$, rather than adopting the category token as the first token in the input token sequence, the strategy adds the category vector to the token vector for feature fusion:

\begin{equation}
\label{variant_1}
    x_i = \omega_i + r + c + p_i,
\end{equation}

where $\omega_i$ is the $i^h$ vector in the input review, $r$ is the rating feature vector, $c$ is the category feature vector, and $p_i$ is the output of positional encoding. The final input sequence to $\rm TRRGen_{\rm Sum}$ is

\begin{equation}
    X = [x_1, x_2, ..., x_n]^T
\end{equation}

Different from $\rm TRRGen_{\rm Sum}$ which sums the feature vector of rating and category with token embedding vector, $\rm TRRGen_{\rm Order}$ regards both category feature vector and rating feature vector as the independent embedding vectors in the input sequence. The embedding vector of tokens in the review is

\begin{equation}
\label{variant_2}
    x_i = \omega_i + p_i,
\end{equation}

where $\omega_i$ is the $i^h$ word embedding vector in the input review and $p_i$ is the output of positional encoding. The final input sequence to $\rm TRRGen_{\rm Order}$ is

\begin{equation}
    X = [c, r, x_1, x_2, ..., x_n]^T
\end{equation}

where $r$ is the rating feature vector and $c$ is the category feature vector.

\begin{table}[t]
\centering
\caption{Comparison results between \tool, $\rm TRRGen_{\rm Sum}$, and $\rm TRRGen_{\rm Order}$ on Dataset A. The best results are highlighted in \textbf{bold} fonts.}
\label{tab:variant}
\begin{tabular}{cccccc}
\toprule
& BLEU-4 & $p_1$ & $p_2$ & $p_3$ & $p_4$  \\
\midrule
$\rm TRRGen_{\rm Sum}$ & 32.11 & 47.56 & 31.39 & 27.52 & 25.88\\
$\rm TRRGen_{\rm Order}$ & 43.78 & 59.88 & 43.86 & 39.45 & 37.09\\
TRRGen & \textbf{45.38} & \textbf{60.81} & \textbf{45.07} & \textbf{40.55} & \textbf{38.15} \\
\bottomrule
\end{tabular}
\end{table}

Table~\ref{tab:variant} shows the comparison results between \tool, $\rm TRRGen_{\rm Sum}$, and $\rm TRRGen_{\rm Order}$ on Dataset A. We can find that the \tool achieves the best performance and $\rm TRRGen_{\rm Order}$ achieves slightly worse performance than \tool. However, the performance of $\rm TRRGen_{\rm Sum}$ signficantly drops and is even worse than the performance of baseline $\rm RRGen$.From the results, we can find that embedding the category information as the first token in the input sequence makes important contributions to performance improvement. Thanks to the long-term dependency learning ability from the self-attention mechanism in the Transformer model, the category token in the input sequence can offer the decoder the category information and guide the decoder to generate suitable responses. On the contrary, the semantic information inside the embedding word vector will be damaged severely if the category embedding vectors are simply added to the embedding token vector. The models will tend to generate the same response for all users' reviews. In addition, the rating vectors offer the emotional information from the users' reviews, and adding such vectors into every token embedding vector can enhance the emotional information hidden in the words, which also contributes to performance improvement.

\section{Threats to Validity}\label{sec:threat}
One of the threats to validity is the limited number of studied apps. The data set we use, which is from \cite{gao2019automating} only contains review-response pairs coming from free apps. The main reason for not using data from non-free apps is that the pricing of an app is likely to impact developers' response behavior \cite{hassan2018studying}. The data set we collect also has the same problem. Another issue is that all the data we study is collected from the apps in Google Play Store. Although our study is based on various kinds of apps and large numbers of review-response pairs, it still can be extended to more apps platforms and paid apps in future work.

The next threat to validity concerns the feature information we use in our model to improve the effectiveness of review response generation. We only utilize the app categories and review ratings associated with user reviews as additional information to help the Transformer model learn and generate the response. Some other review features (e.g., length of review) are also easy to obtain, but we have not found a proper way to use them. In the future, we will explore the influence of more review features on automatic review response generation.

Another threat to validity is the manual inspection in Section \ref{sec:result}. In the survey of human evaluation, we only invite six participants to evaluate the responses. The human evaluation results are also influenced by the participants' experience and their intuition of the evaluation metrics. To reduce the deviation in the manual evaluation, we make sure that three different participants evaluated each review and its responses. As our participants are all students, and they are not likely to benefit from our tools in practice, we ensure that most of them have at least one year of software development experience to alleviate this threat.
\section{Related Work}\label{sec:literature}

\subsection{App Review Mining}
Structural features of review are easy to be obtained for app review analysis, such as review length and TF-IDF \cite{liang2017text}. Content features including sentiment, topic, and keyword are usually extracted by machine learning approaches. Guzman and Maalej \cite{guzman2014users} use the topic modeling approach and StentiStrength \cite{thelwall2010sentiment} (a lexical sentiment extraction tool) to predict sentiment of app features. Identifying the topics expressed by the user reviews is the basis of review mining \cite{palomba2017recommending,grano2018exploring,gao2019emerging}. For example, Di Sorbo et al. \cite{di2016would} separately categorize user intentions and topics delivered by app reviews. The work \cite{bharti2017automatic} also introduces various kinds of methods to extract keywords automatically. A more comprehensive study of app review analysis can be referred to Martin et al.'s work~\cite{DBLP:journals/tse/MartinSJZH17}.


\subsection{Analysis of User-Developer Dialogues in App Stores}
User rating and review function of the app stores serve as a communication channel among users and with developers \cite{pagano2013user}. Users continuously post dozens of comments about user experience, bugs, and feature requests, and developers can also reply to them to improve the user rating. Oh et al. \cite{oh2013facilitating} surveyed 100 smartphone users to understand how developers and users interact. The survey results show that 69\% of users tend to take a passive action such as uninstalling apps because they think their inquiries would take a long time to respond to by developers or receive no response. Harman et al. \cite{harman2012app} found a strong correlation between an app's rating and its download numbers. McIlroy et al. \cite{mcilroy2015worth} observed that users change their rating 38.7\% of the time following response on a data set containing over 10,000 free apps from Google Play Store. Such a fact is also confirmed by Hassan et al. \cite{hassan2018studying}. These studies highlight the importance of responding to a user review, but the researchers did not explicitly provide any automation method to help developers respond. Gao et al. \cite{gao2019automating} are the first to propose to adopt an RNN-based model to encode the review context with features such as rating to respond to the user reviews.

\subsection{Dialogue Generation}
Dialogue Generation is a popular topic in the field of natural language processing. In particular, given a message from a human, the machine automatically generates a relevant reply. An important challenge in such a task is to develop Seq2Seq models that can effectively incorporate dialogue context and generate meaningful and diverse responses \cite{serban2016generative}. Researchers have mainly proposed two types of Seq2Seq models. The first one is usually trained with cross-entropy to generate responses word-by-word conditioned on a dialogue context \cite{li2016deep,li2016diversity,serban2017multiresolution}, called generative models. The second one is trained to select an appropriate response from a knowledge base \cite{lowe2015ubuntu,inaba2016neural,bordes2016learning}, called discriminative models or information retrieval approach. However, both generative and discriminative models are not perfect. The problem with generative models is that they cannot guarantee that the response is a legitimate natural language text. The major bottleneck for discriminative models is the creation of the knowledge base \cite{chen2011evaluating}.
\section{Conclusion and Future Work}\label{sec:conclusion}
In this paper, we propose a novel Transformer-based model named TRRGen for automatic app review response generation. TRRGen is the first Transformer-based model fusing the features of app category and rating. Specifically, TRRGen combines token embedding vectors with rating embedding vectors and app category embedding vectors as the user review embedding. Then TRRGen adopts a Transformer to encode the review embeddings for capturing the review semantics and sentiment. The experimental results demonstrate that TRRGen outperforms the state-of-the-art approaches and the fusion of app category feature and rating feature into token semantics is helpful for generating high-quality replies.

In the future, we will extend our work with larger datasets. We also consider excavating more features from user reviews and applying them into our review response generation model to enhance the task performance.

\section{Acknowledgement}\label{sec:acknowledgement}
This research was supported by National Natural Science Foundation of China Grant under project No. 62002084, Stable support plan for colleges and universities in Shenzhen under project No. GXWD2020 1230155427003-20200730101839009, the Major Key Project of PCL (Grant No. PCL2022A03, PCL2021A02, PCL2021A09), Guangdong Provincial Key Laboratory of Novel Security Intelligence Technologies (2022B1212010005), and the Research Grants Council of the Hong Kong Special Administrative Region, China (CUHK 14210920 of the General Research Fund). 





\bibliography{sigproc}




\end{document}